\newcommand{\fract}[2]{{\textstyle\frac{#1}{#2}}}
\newcommand{\lapeq}{\stackrel{\scriptscriptstyle\raisebox{-3.0mm}{$<$}}
{\scriptscriptstyle \raisebox{-1.5mm}{$\sim$}}}

\documentclass[epj]{svjour}
\usepackage{graphics}
\usepackage{epsfig}
\begin{document}
\title{Flavor Symmetry Breaking and Strangeness in the Nucleon}
\author{H. Weigel
\thanks{\emph{Heisenberg--Fellow}}
}                     
\institute{Institute for Theoretical Physics, T\"ubingen University,
D-72076 T\"ubingen, Germany}
\date{Received: date / Revised version: date}
%
\abstract{We suggest that breaking of $SU(3)$ flavor symmetry mainly
resides in the baryon wave--functions while the charge operators
have no (or only small) explicit symmetry breaking components. We utilize
the collective coordinate approach to chiral soliton models to support
this picture. In particular we compute the $g_A/g_V$ ratios for hyperon 
beta--decay and the strangeness contribution to the nucleon axial current 
matrix elements and analyze their variation with increasing flavor symmetry
breaking.
} 
\maketitle
\section{Introduction and Motivation}

There has been much interest in the strangeness content of the
nucleon ever since the analysis of the DIS data~\cite{EMC} suggested
a large (negative) polarization of strange quarks in the 
nucleon~\cite{Br88,El95}, $\triangle S_N\approx-0.15$. This
surprising result particularly relies on the assumption of
flavor covariance for the axial current matrix elements of the 
$\frac{1}{2}^+$ baryons. This assumption originates from the feature 
that the Cabibbo scheme \cite{Ca63}, that utilizes the $F$\&$D$ 
parameterization for the flavor changing axial charges, works 
unexpectedly well~\cite{Fl98} as the comparison in 
table~\ref{empirical} shows. 

\begin{table}[hb]
~\vskip-0.4cm
\caption{\label{empirical}\sf The empirical values for the 
$g_A/g_V$ ratios of hyperon beta--decays \protect\cite{DATA}, 
see also~\protect\cite{Fl98}. For $\Sigma\to\Lambda$ only $g_A$ is given.
Also the flavor symmetric predictions are presented using
$F=0.459$ and $D=0.799$. Analytic
expressions which relate these parameters to the $g_A/g_V$ ratios
may {\it e.g.} be found in table I of \protect\cite{Pa90}.}
{\small
\begin{tabular}{ c || c | c | c}
& $\Lambda\to p$ & $\Sigma\to n$ & $\Xi\to\Lambda$ \\
\hline
emp.& $0.718\pm0.015$ & $0.340\pm0.017$ & $0.25\pm0.05$ \\
$F$\&$D$& $0.725\pm0.009$ & $0.339\pm0.026$ & $0.19\pm0.02$ \\
\hline
\hline
& $\Xi\to\Sigma$ & $\Sigma\to\Lambda$  \\ 
\hline
emp.& $1.287\pm0.158$ & $0.61\pm 0.02$  \\
$F$\&$D$&$1.258=g_A$ & $0.65\pm0.01$ \\
\end{tabular}
}
\end{table}
\vskip-0.3cm

Here we will investigate in how far this agreement justi\-fies 
to carry over flavor covariance to strangeness conserving axial current
matrix elements in order to disentangle the various quark flavor
components of the nucleon axial current matrix element. This 
investigation requires baryon axial current matrix elements as functions 
of the (effective) strength of flavor symmetry breaking. This can be 
achieved within the three flavor version of the Skyrme model
(and generalizations thereof) in which baryons emer\-ge as solitons.
In such models baryon states are constructed by quantizing 
the large amplitude fluctuations about the soliton and constructing
exact eigenstates in the presence of symmetry breaking. 
We focus on a picture where symmetry breaking 
mainly resides in the baryon wave--functions, 
including important contributions which would be missed in a first
order treatment. In contrast, we assume that the current 
operators, from which the charges are computed, are dominated by 
flavor covariant components. In a first step we do not 
specify the model Lagrangian but adjust the prefactors of
the few possible flavor covariant components of the axial current
operator to observables in hyperon beta--decay and analyze their
matrix elements as functions of flavor symmetry breaking.
We also present results obtained from a 
realistic vector meson soliton model that supports the suggested 
picture. Details omitted here may be traced from ref~\cite{We00}.

\section{Symmetry Breaking in the Baryon Wave--Functions} 

The collective coordinates $A$ that parameterize the large 
amplitude fluctuations off the soliton are introduced via
\begin{equation}
U(\vec{r},t)=A(t)U_0(\vec{r})A^\dagger(t)\, ,\qquad 
A(t)\in SU(3)\, .
\label{collcord}
\end{equation}
$U_0(\vec{r})$ describes the soliton embedded
in the isospin subgroup. A prototype model Lagrangian
for $U(\vec{r},t)$ consists of the Skyrme model 
supplemented by the Wess--Zumino--Witten term and suitable 
symmetry breaking pieces. We parameterize the collective coordinates 
by ``Euler--angles'' 
\begin{equation}
A=D_2(\hat{I})\,{\rm e}^{-i\nu\lambda_4}D_2(\hat{R})\,
{\rm e}^{-i(\rho/\sqrt{3})\lambda_8}\ .
\label{Apara}
\end{equation}
Here $D_2$ denote the rotation matrices for rotations in 
isospace~($\hat{I}$) and coordinate--space~($\hat{R}$). 
Substituting (\ref{collcord}) into the model Lagrangian yields 
upon canonical quantization the Hamiltonian for the
collective coordinates~$A$: 
\begin{equation} 
H=H_{\rm s}+\fract{3}{4}\, \gamma\, {\rm sin}^2\nu\, .
\label{Hskyrme}
\end{equation}
The symmetric piece of this collective Hamiltonian only contains 
Casimir operators and may be expressed in terms of the $SU(3)$--right 
generators $R_a\, (a=1,\ldots,8)$:
\begin{equation}
H_{\rm s}=M_{\rm cl}+\frac{1}{2\alpha^2}\sum_{i=1}^3 R_i^2
+\frac{1}{2\beta^2}\sum_{\alpha=4}^7 R_\alpha^2\, .
\label{Hsym}
\end{equation}
$M_{\rm cl},\alpha^2,\beta^2$ and $\gamma$ are functionals of the 
soliton, $U_0(\vec{r})$. The generators $R_a$ can be expressed in terms 
of derivatives with respect to the `Euler--angles'. The essential 
feature of the parameterization~(\ref{Apara}) is that the flavor 
symmetry breaking part of the full Hamiltonian~(\ref{Hskyrme}) only
depends on the flavor changing angle~$\nu$. Therefore the eigenvalue 
problem $H\Psi=\epsilon\Psi$ reduces to ordinary second order 
differential equations for isoscalar functions which only depend on 
$\nu$~\cite{Ya88}. Solely the product $\omega^2=\frac{3}{2}\gamma\beta^2$ 
appears in these differential equations as the effective strength of 
symmetry breaking on which the eigenfunctions of $H$ depend 
parametrically. A value in the range $5\lapeq\omega^2\lapeq8$ 
is required to obtain reasonable agreement with the empirical mass 
differences for the $\frac{1}{2}^+$ and $\frac{3}{2}^+$ 
baryons~\cite{We96}. Such large a value for $\omega^2$ is without 
reach of a perturbation expansion as the resulting baryon wave--functions
exhibit strong distortion from flavor covariance.

\section{Charge Operators} 

In the soliton description the effect of the derivative type symmetry 
breaking terms is mainly indirect. They provide the splitting between the 
various decay constants and thus increase $\gamma$ since it is proportional 
to $f_K^2m_K^2-f_\pi^2m_\pi^2\approx 1.5f_\pi^2(m_K^2-m_\pi^2)$. 
Otherwise the derivative type symmetry breaking terms are negligible. 
Whence symmetry breaking terms can be omitted in the current operators 
and the non--singlet axial charge operator is parameterized as 
($ a=1,\ldots,8,\, i=1,2,3$)
\begin{equation}
\int d^3r A_i^{(a)} = c_1 D_{ai} - c_2 D_{a8}R_i
+c_3\sum_{\alpha,\beta=4}^7d_{i\alpha\beta}D_{a\alpha}R_\beta
\, ,
\label{axsym}
\end{equation}
where
$D_{ab}=\frac{1}{2}{\rm tr}\left(\lambda_a A\lambda_b A^\dagger\right)$.
For $\omega^2\to\infty$ (infinitely heavy {\it strange} degrees of freedom)
the strangeness contribution to the nucleon axial charge should vanish. 
Noting that
$\langle N| D_{83}| N\rangle\to0$ and 
$\langle N| \sum_{\alpha,\beta=4}^7d_{3\alpha\beta}D_{8\alpha}R_\beta
| N\rangle\to0$ while $\langle N| D_{88}| N\rangle\to1$ for 
$\omega^2\to\infty$, we demand 
\begin{equation}
\int d^3r A_i^{(0)}= -2\sqrt{3} c_2 R_i\quad i=1,2,3
\label{singsym}
\end{equation}
for the axial singlet current because it leads to the strange\-ness
projection, $A_i^{(s)}=(A_i^{(0)}-2\sqrt{3}A_i^{(8)})/3$ that vanishes
for $\omega^2\to\infty$.
Actually all model calculations in the literature \cite{Pa92,Bl93} 
are consistent with this relation between singlet and octet currents.
The singlet current matrix element, $\triangle\Sigma_B=\sqrt3 c_2$, is
the quark spin contribution to the spin of the considered baryon, $B$. It
is well known that the empirical value for the nucleon matrix element,
$\triangle\Sigma_N\approx 0.20\pm0.10$~\cite{El95} is insensitive to the
strength of flavor symmetry breaking~\cite{Jo90}. This suggests to
adjust $c_2$ accordingly. In order to completely describe the hyperon 
beta--decays we also demand matrix elements of the vector charges. These 
are obtained from the operator
\begin{equation}
\int d^3r V_0^{(a)} = \sum_{b=1}^8D_{ab}R_b=L_a,
\label{vector}
\end{equation}
which introduces the $SU(3)$--left generators $L_a$. 

\begin{figure}[t]
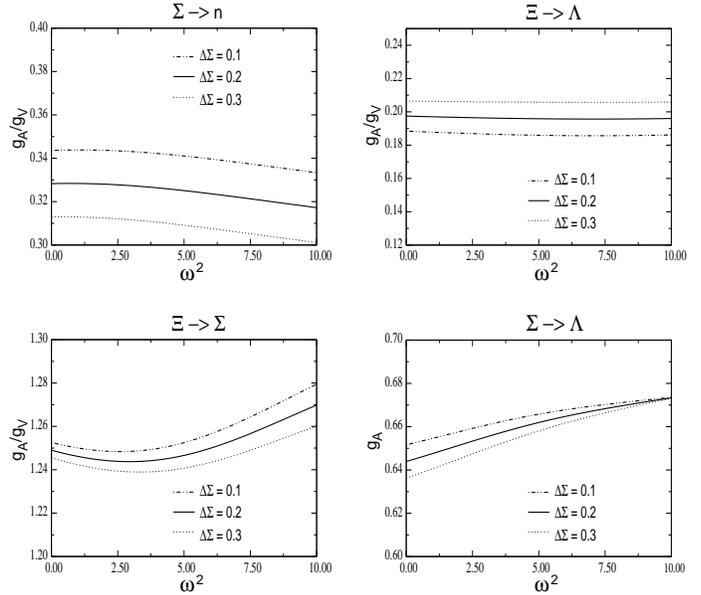

\centerline{
\epsfig{figure=nusi.eps,height=4.0cm,width=3.7cm,angle=270}
\hspace{0.5cm}
\epsfig{figure=chla.eps,height=4.0cm,width=3.7cm,angle=270}}
~\vskip0.03cm
\centerline{
\epsfig{figure=sich.eps,height=4.0cm,width=3.7cm,angle=270}
\hspace{0.5cm}
\epsfig{figure=sila.eps,height=4.0cm,width=3.7cm,angle=270}}
\caption{\label{decay}\sf The predicted decay parameters for the 
hyperon beta--decays using $\omega^2_{\rm fix}=6.0$. 
The errors originating from those in $\Delta\Sigma_N$ are indicated.}
\end{figure}

The values for $g_A$ and $g_V$ 
(only $g_A$ for $\Sigma^+\to\Lambda e^+\nu_e$)
are obtained from the matrix elements of the operators in 
eqs~(\ref{axsym}) and~(\ref{vector}), respectively, sandwiched 
between the eigenstates of the full Hamiltonian~(\ref{Hskyrme}).
We still have to specify $c_1$ and $c_3$. We determine these two
parameters such that nucleon axial charge, $g_A$ and the $g_A/g_V$ 
ratio for $\Lambda\to p e^-\bar{\nu}_e$ are reproduced\footnote{In 
this section we will not address the problem of the too small model 
prediction for $g_A$.} at a prescribed strength of flavor symmetry
breaking, $\omega^2_{\rm fix}=6.0$. Then we are not only left with 
predictions for the other decay parameters but we can in particular 
study the variation with symmetry breaking. This is shown in 
figure~\ref{decay}.
The dependence on flavor symmetry breaking is very 
moderate\footnote{However, the individual matrix elements
entering the ratios $g_A/g_V$ vary strongly with 
$\omega^2$~\cite{We00}.} and the results can be viewed as reasonably 
agreeing with the empirical data, {\it cf.} table \ref{empirical}. 
The observed independence of $\omega^2$ shows that these
predictions are not sensitive to the choice of $\omega^2_{\rm fix}$.
The two transitions, $n\to p$ and $\Lambda\to p$, which are not shown in
figure~\ref{decay}, exhibit a similar negligible dependence on $\omega^2$.
We therefore have a two parameter ($c_1$ and $c_3$, $c_2$ is fixed 
from $\Delta\Sigma_N$) fit of the hyperon beta--decays. Comparing the 
results in figure \ref{decay} with the data in table~\ref{empirical} 
we see that the present calculation using the strongly 
distorted wave--functions agrees equally well with the empirical 
data as the flavor symmetric $F$\&$D$ fit. On the other hand, the
strangeness contribution to the nucleon axial current matrix element
reduces from $\triangle S_N\approx-0.13$ in the symmetric treatment
to $\triangle S_N\approx-0.07$  in the realistic case.

\section{Model Calculation}

We consider a realistic soliton model containing pseudo\-scalar 
and vector meson fields. It has been established for two flavors in 
ref \cite{Ja88} and been extended to three flavors in ref~\cite{Pa92} 
where it has been shown to fairly well describe the parameters of hyperon 
beta--decay ({\it cf.} table~4 in ref~\cite{Pa92}). 
The model Lagrangian contains terms which 
involve the Levi--Cevita tensor $\epsilon_{\mu\nu\rho\sigma}$, to 
accommodate processes like $\omega\rightarrow3\pi$ \cite{Ka84}. 
Such terms contribute to $c_2$ and $c_3$.  A minimal set of 
symmetry breaking terms is included \cite{Ja89} to account 
for different masses and decay constants. They add 
symmetry breaking pieces to the axial charge operator,
\begin{eqnarray}
\delta A_i^{(a)}&=&c_4 D_{a8}D_{8i}+
c_5 \sum_{\alpha,\beta=4}^7d_{i\alpha\beta}D_{a\alpha}D_{8\beta}
\nonumber \\ && 
+c_6 D_{ai}(D_{88}-1)\,,
\nonumber \\
\delta A_i^{(0)}&=& 2\sqrt{3}\,c_4D_{8i}\, .
\nonumber
\end{eqnarray}
Unfortunately the model parameters cannot be completely 
determined in the meson sector~\cite{Ja88}. We use the remaining 
freedom to accommodate baryon properties in three different ways as
shown in table \ref{realistic}. The set 
denoted by `b.f.' refers to a best fit to the baryon spectrum.
It predicts the axial charge somewhat on the low side, $g_A=0.88$.
The entry `mag.mom.' labels parameters that yield
magnetic moments close to the empirical data
(with $g_A=0.98$) and finally the set labeled `$g_A$' reproduces the 
axial charge of the nucleon~\cite{Pa92}.
\begin{table}[b]
\caption{\label{realistic}\sf Quark spin content of the nucleon 
and the $\Lambda$ in the realistic vector meson model. 
Three sets of model parameters are considered, see text.}
{\small
\begin{tabular}{ c || c | c |c |c}
& \multicolumn{4}{c}{$N$}\\
\hline
& $\Delta U$ & $\Delta D$ & $\Delta S$ & $\Delta\Sigma$\\
\hline
b.f. &
$0.603$&$-0.279$&$-0.034$&$0.291$\\
mag. mom. &
$0.636$&$-0.341$&$-0.030$&$0.265$\\
$g_A$ &
$0.748$&$-0.476$&$-0.016$&$0.256$\\
\hline\hline
& \multicolumn{3}{c}{$\Lambda$} & \\
\hline
&$\Delta U = \Delta D$ & $\Delta S$ & $\Delta\Sigma$ & \\
\hline
b.f.
&$-0.155$&$0.567$&$0.256$&\\
mag. mom.
&$-0.166$&$0.570$&$0.238$&\\
$g_A$
&$-0.164$&$0.562$&$0.233$&\\

\end{tabular}}
\end{table}
We observe that in particular the strangeness projection of the nucleon
axial current is very small and depends only mildly on the model parameters.
This confirms the above conclusion from the general structure of the 
axial current matrix elements that the strangeness admixture in the 
nucleon is significantly smaller than an analysis based on flavor 
covariance suggests. Also the predictions for the axial properties
of the $\Lambda$ hyperon are quite insensitive to the model parameters.
Sizable polarizations of the {\it up} and {\it down} quarks in the
$\Lambda$ are predicted; comparable to those obtained from the 
SU(3) analysis~\cite{Ja96} of the available data.
\section{Conclusions}

We have suggested a picture for the axial charges of the 
low--lying~$\frac{1}{2}^+$~baryons which manages to reasonably reproduce 
the empirical data without introducing (significant) flavor symmetry 
breaking components in the corresponding operators. Rather, a sizable 
symmetry breaking, as demanded by the baryon spectrum, resides almost 
completely in the baryon wave--functions. In this picture the empirical 
data for hyperon beta--decay are as reasonably reproduced as in the 
Cabibbo scheme. We emphasize that the present picture is not a 
re--application of the Cabibbo scheme since here
the `octet' baryon wave--functions have significant 
admixture of higher dimensional representations. Especially, when compared
with the flavor covariant treatment, the present approach predicts a
sizable suppression of strangeness in the nucleon.

\bigskip
{\small 
This work is supported by the Deutsche Forschungsgemeinschaft (DFG) 
under contract We 1254/3-2.
}

\end {document}